\documentstyle[prl,aps]{revtex}

\def\ll{\label}
\def\re{\ref}
\def\c{\cite}

\def\r1{(\ref{$1})}
\def\ot{\otimes}

\def\th{\theta}
\def\ba{\begin{array}{c}}

\def\ea{\end{array}}
\def\pr{\prod}

\def\si{\sigma}
\def\da{\dagger}
\def\De{\Delta}

\def\bet{\beta}
\def\ov{\over}
\def\ha{{1\over 2}}

\def\l{\left}
\def\l({\left(}
\def\r){\right)}
\def\r{\right}
\def\rw{\rightarrow}

\def\la{\lambda}
\def\al{\alpha}
\def\be{\begin{equation}}
\def\bc{\begin{center}}
\def\ec{\end{center}}
\def\ee{\end{equation}}
\def\ed{\end{document}}
\def\bea{\begin{eqnarray}}
\def\eea{\end{eqnarray}}

\begin{document} 
\title{Algebraic approach in unifying quantum integrable models}

\author{
Anjan Kundu \footnote {email: anjan@tnp.saha.ernet.in} \\  
  Saha Institute of Nuclear Physics,  
 Theory Group \\
 1/AF Bidhan Nagar, Calcutta 700 064, India.
 }
\maketitle
\vskip 1 cm

\begin{abstract} 
A novel   algebra underlying  integrable systems is shown to generate and
 unify a large class of quantum integrable models with given $R$-matrix,
 through reductions of an
 ancestor Lax operator and its  different realizations.  Along with known
  discrete and field models a new class of inhomogeneous
and impurity models
 are obtained.

\medskip
 PACS numbers  03.65.Fd,  02.20.Sv, 05.50+q, 11.10.Lm
\end{abstract}

\smallskip

 The self-dual Yang-Mills equation with possible reductions  has
given a vivid unifying picture in classical integrable systems in $1+1$ and
$0+1$ dimensions \c{sdym}. However, in the quantum case not much has been
achieved in this direction and there exists a
 genuine need for discovering some
 scheme, which would generate models of quantum
 integrable systems (QIS) \c{kulskly} along with their
 Lax operators and  $R$-matrix and thus unify them.

The significance of algebraic structures in describing physical consequences
is well recognized. Like  Lie algebras, their quantum deformations
\c{drinfeld} also found  to be of immense importance in physical models
\c{qoptics}-\c{korepin}. In fact the idea of quantum
Lie algebra, which  attracted enormous interest in recent years
\c{chari}-\c{frt} has stemmed from the  QIS and at the same time
has made profound influence on the QIS itself
\c{tarasov}-\c{saclay}.

Motivated by these facts and our experience \c{construct}, we   find  a
novel Hopf algebra as a consequence of the
 integrability condition, which underlies integrable
  models with $ 2 \times 2 $ Lax operators and the trigonometric
  $R$-matrix.  This is more general than the
  well known quantum Lie algebra and in contrast represents a deformed
 quadratic algebra (QdA), so called due to the appearance of generators in
  quadratic form in the defining  algebraic relations.  At the same time it
 unifies a large class of quantum integrable models by generating them in a
 systematic way through reductions of an { ancestor} model with explicit Lax
 operator realization. Note that the Lax operator together with the quantum
$R$-matrix define an integrable system completely, giving also all conserved
quantities including the Hamiltonian of the model.

The   proposed algebra may be given by the simple relations 
\be
 [S^3,S^{\pm}] = \pm S^{\pm} , \ \ \ [ S^ {+}, S^{-} ] =
 \left ( M^+\sin (2 \al S^3) + {M^- } \cos
( 2 \al S^3  ) \right){1 \over \sin \al}, \quad  [M^\pm, \cdot]=0,
\ll{nlslq2a}\ee
where  $ M^\pm$ are  the central elements.  
 We show that (\ref{nlslq2a}) is  not merely a { modification} of known 
$U_q(su(2))$, but   
  a QdA underlying an integrable ancestor model and in effect is dictated
  by the quantum Yang Baxter equation (QYBE) \c{dvega0} $RL \tilde
L= \tilde L L R$. The associated quantum
  $R(\la)$-matrix is the known trigonometric solution 
related to  sine-Gordon (SG)    \c{kulskly},
while the   
   Lax operator may be taken  as
\be
L_t^{(anc)}{(\xi)} = \left( \begin{array}{c}
  \xi{c_1^+} e^{i \al S^3}+ \xi^{-1}{c_1^-}  e^{-i \al S^3}\qquad \ \ 
2 \sin \al  S^-   \\
    \quad  
2 \sin \al  S^+    \qquad \ \  \xi{c_2^+}e^{-i \al S^3}+ 
\xi^{-1}{c_2^-}e^{i \al S^3}
          \end{array}   \right), \quad
          \xi=e^{i \alpha \la}. \ll{nlslq2} \ee
with $c^{\pm}_a$ central to (\re{nlslq2a}) 
 relating  $ M^\pm=\pm   \sqrt {\pm 1} ( c^+_1c^-_2 \pm
c^-_1c^+_2 ). $ 
 The derivation of
algebra (\ref{nlslq2a}) follows from QYBE by inserting  the explicit form 
  (\ref{nlslq2}) and the $R$-matrix   and matching different powers of the
spectral parameter $\xi$.

Note that  (\ref{nlslq2a})  is   
a Hopf algebra \c{hopf} and a generalization of  $U_q(su(2)).$ However,
 unlike Lie algebras or their deformations, due to the presence of
multiplicative operators $M^\pm,$ (\ref{nlslq2a}) becomes
quantum-deformation of a QdA. Since these operators have arbitrary
eigenvalues including zeros, they can not be removed by scaling 
and therefore generically (\ref{nlslq2a}) is different from the known
quantum algebra. Moreover different representations of $M'$s generate new
structure constants leading to a rich variety of deformed Lie algebras,
which are
related to different integrable systems. This fact becomes important for its
present application.
 The appearance of QdA in basic integrable system should be rather  expected,
since the QYBE with $R$-matrix having $c$-number elements is itself a QdA.
 The notion of QdA  was  introduced first by Sklyanin
\c{sklyalg}.

The  { ancestor model} can be
constructed through representation of (\ref{nlslq2a}) in physical variables
( with
 $  [u,p]=i,$)  as \c{g}
\be
 S^3=u, \ \ \   S^+=  e^{-i p}g(u),\ \ \ 
 S^-=  g(u)e^{i p}.
\ll{ilsg}\ee
  This gives a  novel  exactly integrable quantum
 system {\it generalizing lattice  SG model} and 
 associated with the Lax operator (\ref{nlslq2}).
It is evident that for hermitian $g(u)$ only one gets $S^-=(S^+)^\dag$.
We show below
 that through  various realizations of the single object
$g(u)$  in  (\ref{ilsg}) the ancestor model generates a whole class of
   integrable models. Their Lax operators are derived
 from (\ref{nlslq2}), while the
$R$-matrices are  simply inherited.  The underlying algebras are given by
the corresponding representations of (\ref{nlslq2a}).

 Evidently, fixing 
  $ M^-=0, M^+=1. $ (\ref{nlslq2a})
 leads to the well known quantum algebra
 $U_q(su(2))$ \c{jimbo}. Now the simplest representation ${\vec S} = 
  {\vec \si}$ derives  the integrable $XXZ$ {\it spin chain} \c{fad95},
while
  (\ref{ilsg}) with the corresponding reduction of  $g(u)$ \c{g}
 yields
  the {\it lattice sine-Gordon} model \c{korepinsg} with
its  Lax operator  obtained  from (\ref{nlslq2}) with all $c'$s $=1$.

An asymmetric   choice of  central elements:
$ \ \ c^+_{1,2}=1, \ c^-_1= {1 \ov c^-_2}= -{iq }, q=e^{i \al}  $
  along with  the  mapping 
$ S^+= c A, \ S^-= c A^\dag, \ S^3= -N, \ \
 c=  (\cot \al)^\ha
$
 brings (\ref{nlslq2a}) directly
 to the well known
$q$-oscillator algebra \c{qoscl,ng} and  
  simplifies $g^2(u)=  
{[-2u]_q}$ in  (\ref{ilsg}). 
 Therefore using the 
inter-bosonic  map \c{map}  one gets 
 a bosonic realization for the $q$-oscillator \c{map}. This realization
in turn constructs easily from (\ref{nlslq2})
the Lax operator, which coincides exactly with 
 the  discrete version of the quantum {\it derivative nonlinear
Schr\"odinger equation} (QDNLS) \c{construct}. The QDNLS
 was shown to be  related to the interacting bose gas with derivative
$\delta$-function potential
\c{shirman}. Fusing   two such models one can further create an integrable
 {\it massive Thirring}
 model described in \c{kulskly}. 

 Having the freedom of choosing  trivial eigenvalues for the central elements:
$  \ c^-_1=c^+_2=0 $  with other $c'$s$=1$  we  obtain another  
 deformed  Lie
 algebra $ \ \
[ S^+, S^-]= {e^{2i\al S^3} \over  i \sin \al }.$
 This can be  realized  again by  (\ref{ilsg}) with the
 related
expression for $g(u)$ \c{g}, 
  using which   the   Lax operator is obtained  from (\ref{nlslq2}).
 The  model thus results 
is no other than  the  {\it discrete 
quantum Liouville} model \c{llm}.
 Note that  the present case: $ M^\pm=\pm  \sqrt {\pm 1}$
  may be achieved even with 
$ \ c^-_1 \not =0$ giving  
 the same  algebra and hence the same realization. However, the Lax operator
which depends explicitly on $c$'s gets changed reducing 
(\ref{nlslq2}) to another nontrivial structure. This is an interesting
possibility of constructing different useful Lax operators for the same
model, in a systematic way. For example, the present construction of
 the {\it second Liouville Lax}
operator recovers that of  
\c{fadliu}, invented 
 for its Bethe ansatz solution.

 In a similar way the particular case $M^\pm=0$ can be achieved with
  different sets of choices: with all  $c'$s $=0$  except  $ 
i)  c_a^+=1 \ , \
 ii)  c_1^\mp=\pm 1  , \  iii)  c_1^+=1, \ $ 
 all of which lead to the same   algebra
\be
[ S^+, S^-]= 0, \ [ S^3, S^\pm]= \pm S^\pm .\ll{nula} \ee
However, they may generate   different   
  Lax operators  from (\ref{nlslq2}), 
which  might  even  correspond to different  models, though 
with  the same underlying algebra.  In particular, case i) leads to
  the {\it light-cone SG} model, while
 ii) and iii)   give two different  Lax operators  found in
\c{rtoda} and \c{hikami} for  the same  { relativistic 
 Toda chain}.
Since  here we get
  $g(u)=$const. ,  interchanging $u \rw -ip, p\rw -i u,$
  (\ref{ilsg}) yields  simply
$ \ \
 S^3 =-ip, \ S^\pm=  \al e^{\mp u } \ \ $
  generating     {\it discrete-time or relativistic
quantum Toda chain}.

 Remarkably,  all the descendant models  listed above
have the same trigonometric $R$-matrix 
 inherited from the ancestor model and similar is true
  for   its rational form, as we will see below.  This unveils the
mystery why a wide range  of  models found to share the same
$R$-matrices.  The $L$-operators 
and the underlying algebras however 
become different, being various reductions of the ancestor Lax operator 
   (\ref{nlslq2})  and the
ancestor algebra (\ref{nlslq2a}).

We  consider now  the  undeformed   $\al \rw 0$ limit   
  of the proposed algebra  (\ref{nlslq2a}). It is evident that for  
 the  limits to be finite the central elements 
 must also be $\al $ dependent.
A consistent procedure leads to $S^\pm \rw  is^\pm, 
M^+ \rw -m^+, M^- \rw -\al
m^-, \xi \rw 1+ i \alpha \la $ giving the  algebraic relations 
\be  [ s^+ , s^- ]
=  2m^+ s^3 +m^-,\ \ \ \ 
  ~ [s^3, s^\pm]  = \pm s^\pm 
  \ll{k-alg} \ee
with $m^+=c_1^0c_2^0, m^-= c_1^1c_2^0+c_1^0c_2^1$ as
the new central elements.  It is again not a Lie but a
 QdA, since  multiplicative   operators
  $m^\pm$ can not be removed in general due their  allowed  zero
eigenvalues.  (\ref{k-alg}) exhibits also
 noncocommutative feature \c{hopf} unusual for an undeformed algebra. 
  (\ref{nlslq2}) at
this limit reduces to
 \be
L_r{(\la)} = \left( \begin{array}{c}
 {c_1^0} (\la + {s^3})+ {c_1^1} \ \ \quad 
  s^-   \\
    \quad  
s^+    \quad \ \ 
c_2^0 (\la - {s^3})- {c_2^1}
          \end{array}   \right), \ll{LK} \ee
 while $R$-matrix is converted into its rational form, well known for the
NLS model \c{kulskly}.
Therefore the integrable systems associated with  algebra (\ref{k-alg}) and
generated by ancestor model (\ref{LK})
 would belong to  the rational class all sharing the same rational
$R$-matrix.

It is interesting to find that   the bosonic representation
 (\ref{ilsg}), using the undeformed limit $g_0(u)$ \c{g} and the
 inter-bosonic map \c{map}
  reduces into a generalized Holstein-Primakov transformation (HPT)
 \be
 s^3=s-N, \ \ \ \    s^+= g_0(N) \psi, \ \ \  s^-= \psi^\dag g_0(N)
, \ \ \ \ g_0^2(N)=m^-+m^+ (2s -N), \ \ N=\psi^\dag \psi.
\ll{ilnls} \ee
It can be checked  to be  an exact realization of  (\ref{k-alg}),
associated with the Lax operator (\ref{LK}).
 This
  would serve therefore as an  ancestor model of rational class and
represent an  integrable {\it generalized  lattice NLS} model.

For the choice 
 $  m^+  =  1,m^-  =  0, $
  (\ref{k-alg}) leads  clearly to the standard
   $su(2)$ and for  spin $\ha$
representation recovers the  $XXX$ {\it  spin chain}\c{dvega0}.
On the other hand the general form (\ref{ilnls}) simplifies 
 to  standard  HPT and (\ref{LK}) reproduces
the {\it lattice   NLS} model 
\c{korepinsg}.
 
The complementary  choice $  m^+  =  0,m^-  =  1, $ reduces (\ref{k-alg}) to
    a non semi-simple algebra
and gives  $g_0(N)=1.$ This induces a direct 
realization through oscillator algebra:
$s^ {+}=\psi, s^ {-}=\psi^\dag, s^ {3} = s -N $
and  corresponds to another {\it simple   
lattice NLS} model
\c{kunrag}. 
Remarkably, further trivial choice $m^-  = 0$  gives again 
  algebra (\ref{nula}) and   therefore the same
  realization found before for the relativistic case can  be used, but  
   now  for the   {\it nonrelativistic
 Toda chain} \c{kulskly}. The associated Lax operator 
 should  however  be obtained   from (\ref{LK}) along with    the 
 rational  $R$-matrix.

It should be noted that a bosonic realization  of 
  general Lax operators like (\ref{nlslq2}) and
({\ref{LK}) can be found also   in  some earlier works  \c{tarasov,korl}.
Apart from the discrete models obtained above, one can 
 construct a family of   quantum field models starting 
    from their lattice versions. Scaling first  the operators like
$ p_j, u_j , c^\pm_a, 
 \psi_j,$ 
 consistently  by lattice spacing $\Delta$ and taking  the
continuum limit
$\Delta \rw 0$ one gets $p_j \rw p(x), \psi_j \rw \psi(x)$ etc.  
The Lax operator ${\cal L}(x, \la)$ for the continuum model then obtained
from its discrete counterpart as $L_j(\la) \rw I+ i\De {\cal L}(x).$ The
associated $R$-matrix however remains the same since it does not contain
$\De$. Thus {\it integrable field models} like sine-Gordon,
 Liouville, NLS or the derivative NLS models are obtained from their
discrete variants constructed above.


It is  possible  to   build further a 
  new class
of  models, that may be considered as the 
  inhomogeneous versions  of the  above integrable models. The idea of  such
 construction is to take locally different representations for the central
elements, i.e. instead of taking their  fixed eigenvalues one
 should consider them to be site dependent functions.
This simply means that
 in the expressions of $g(u_j)$ \c{g}, $M^\pm$ should be replaced
by  $ M^\pm_j$ and consequently   in  Lax operator
(\ref{nlslq2})  all $c'$s should be changed to $ c_{j}'$s. 
Thus in  lattice models the values of  central elements
may vary arbitrarily   at  different
lattice points $j$  including zeros. This would naturally 
lead to  inhomogeneous lattice models. However since the algebra 
remains the same they  answer to the same quantum
$R$-matrices. Physically such inhomogeneities may be interpreted as 
impurities, varying  external fields,  incommensuration etc.

   Notice that in the sine-Gordon model unlike its coupling constant the
mass parameter enters through the Casimir operator of the underlying
algebra. Therefore taking
$ M_j^+=- (\De m_j)^2,$ one can construct a variable mass discrete
SG model without spoiling its integrability.
 In the continuum field limit it would generate a novel 
  {\it sine-Gordon 
 model with variable mass} $m(x)$ in an  external gauge field $\th (x)$.
In the simplest case the Hamiltonian of such model would be
${\cal H}= \int dx \left ( m(x) (u_t)^2 + m^{-1}(x) (u_x)^2 + 8(m_0-m(x)
\cos (2 \al u )) \right). $ Similar models
 may arise also in physical situations
\c{msg}. 

{\it Inhomogeneous lattice NLS} model can   be obtained by
considering site-dependent values for central elements in (\ref{LK}) and in
the
  generalized HPT (\ref{ilnls}), where time dependence can also enter as
parameter. As a possible quantum field model it would correspond to
equations like {\it cylindrical NLS}
 \c{rl} with explicit coordinate dependent coefficients. In a similar way
 inhomogeneous versions of Liouville model, relativistic 
Toda  etc. can be constructed. For example, taking 
$ c^a_{1} \rw c^a_{j} $ in  nonrelativistic Toda chain  we can get  a new   integrable
 quantum {\it Toda chain with inhomogeneity}
 having the  Hamiltonian
$ \ \ H= \sum_j (p_j +{c^1_j \ov
 c^0_j})^2+{1 \ov c^0_jc^0_{j+1}} e^{u_j-u_{j+1}}. \ $

Another way of 
constructing  inhomogeneous models 
 is to  use different realizations of
 the  general QdA (\ref{nlslq2a}) or (\ref{k-alg}) at different
lattice sites,
depending on the type of  $R$-matrix. This may lead even to different
underlying algebras and hence different Lax operators at differing sites
opening up possibilities of building various exotic inhomogeneous
integrable models.  Thus in a simple example of impurity   $XXX$ spin
chain if we replace its standard Lax operator at a single impurity 
site $m$  by a
compatible $L_{am}= (\la+ c_m^1)\si^3_a,$
 the Hamiltonian 
 of the model
 is modified to $
H=- \left( \sum_{j \not = m,m-1}\vec \si_j \vec \si_{j+1} +h_{m-1m+1}
\right ), $ where $h_{m-1m+1}=-
(\si^+_{m-1}\si^-_{m+1} + \si^-_{m-1}\si^+_{m+1})+ 
\si^3_{m-1}\si^3_{m+1} .$
It gives an   integrable quantum spin chain {\it with a defect}, where
the coupling constant has changed sign at the impurity site. If attempt is
made to restore the sign it appears in the boundary condition.



Thus we have    prescribed  an unifying scheme for 
 quantum
integrable systems, where the models can be   generated systematically
 from a single ancestor model with  underlying algebra
(\ref{nlslq2a}).  
 The Lax operators of the descendant  models are constructed 
from (\ref{nlslq2}) or its $q \rw  1 $ limit (\ref{LK}), while the 
variety of their concrete representations 
  are  obtained from the same  
 general form (\ref{ilsg})  at different realizations.
 The corresponding underlying algebraic structures
are the allowed reductions of (\ref{nlslq2a}). The associated quantum
$R$-matrix  however remains the same
   trigonometric  or the rational form  as
inherited from the ancestor model. This fact also reveals a universal 
character for solving the models through
 algebraic Bethe ansatz (ABA) \c{kulskly,fadrev}.
 The characteristic  eigenvalue equation for the ABA is given by    
$ \Lambda_m (\la) =
\al(\la) \pr_{j=1}^m {f(\la_j-\la)}  
+\bet(\la)
\pr_{j=1}^m {f(\la-\la_j)}$ 
, 
where the coefficients $\al (\la)$ and $\bet (\la) $ are the only
model-dependent elements, as being eigenvalues of the pseudovacuum they
 depend on the concrete form 
of the Lax operator.
 The  main bulk of the  expression however is  given through functions like
$ \ {f(\la) }= {a(\la) \ov b(\la)},$  i.e. 
 as the ratio  of two elements  of the $R$-matrix and hence is
universal for all models belonging to the same class. Therefore all
integrable models solvable through ABA can be given by almost
 a universal 
 equation based on a general model.

\medskip

 The author  thanks Prof. Y. Ng 
for his kind hospitality and encouragement at  N. Carolina Univ.
where this work was completed and Prof. Alexios Polychronakos for fruitful
comments.

 \end{document}